\documentclass[epj]{svjour}
\usepackage{graphics}
\usepackage{epsfig}
 \usepackage{graphicx}
 \usepackage{color}
 \usepackage{mathrsfs} 
 \usepackage{lscape,epsfig}
 \usepackage{amsmath}
 \usepackage{bm}
 \usepackage{txfonts}
 \usepackage{multirow}
 
  \def\beq{\begin{equation}}
  \def\eeq{\end{equation}}
  \def\beqa{\begin{eqnarray}}
  \def\eeqa{\end{eqnarray}}
  \def\ban{\begin{eqnarray*}}
  \def\ean{\end{eqnarray*}}
  \def\bi{\begin{itemize}}
  \def\ei{\end{itemize}} 
  
\begin{document}

\title{Impact of chiral hyperonic three-body forces on neutron stars}

\author{Domenico Logoteta  \inst{1,2} \and Isaac Vida\~na \inst{3} \and Ignazio Bombaci \inst{1,2}}  %
\institute{Dipartimento di Fisica ``Enrico Fermi'', Universit\'a di Pisa, Largo Pontecorvo 3, 56127 Pisa, Italy \and 
Istituto Nazionale di Fisica Nucleare, Largo Pontecorvo 3, 56127 Pisa, Italy \and 
INFN, Sezione di Catania, Dipartimento di Fisica ``Ettore Majorana", Universit\`a di Catania, Via Santa Sofia 64, I-95123 Catania, Italy}

\date{Received: date / Revised version: date}

\abstract{
We study the effects of the nucleon-nucleon-lambda (NN$\Lambda$) three-body force on neutron stars. In particular, we consider the NN$\Lambda$ force recently derived 
by the J\"{u}lich--Bonn--Munich group within the framework of chiral effective field theory at next-to-next-to-leading order. This force,  together with realistic nucleon-nucleon, nucleon-nucleon-nucleon and 
nucleon-hyperon interactions, is used to calculate the equation of state and the structure of neutron stars within the many-body non-relativistic Brueckner-Hartree-Fock approach.
Our results show that the inclusion of the NN$\Lambda$ force  leads to an equation of state stiff enough such that the resulting neutron star maximum mass is compatible with the largest currently measured ($\sim 2\ M_\odot$) neutron star masses. 
Using a perturbative many-body approach we calculate also the separation energy of the $\Lambda$ in some hypernuclei
finding that the agreement with the experimental data improves for the heavier ones when the effect of the NN$\Lambda$ force is taken into account.
\PACS{
	  {97.60.Jd}{Neutron stars} \and
    {04.30.Tv}{Gravitational-wave astrophysics} 
     } 
} 

\authorrunning{Domenico Logoteta, Isaac Vida\~na and Ignazio Bombaci}
\titlerunning{Impact of chiral hyperonic three-body forces on neutron stars}
\maketitle

\section{Introduction}

The importance of taking into account nucleon-nucleon-nucleon (NNN) interactions in finite nuclei as well as in infinite nuclear matter is 
nowadays a well established feature. 
It is well known that high precision nucleon-nucleon (NN) potentials, which fit NN scattering 
data up to an energy of $350$ ${\rm MeV}$ with a $\chi^2$ per datum close to 1, underestimate the experimental 
binding energies of $^3$H and $^3$He by about $1$~${\rm MeV}$, and that of $^4$He by about $4$ ${\rm MeV}$\ \cite{kievsky2008}. 
This missing binding energy can be accounted for by introducing a NNN interaction into the nuclear Hamiltonian~\cite{kievsky2008}. 
Three-nucleon forces are also crucial for nuclear matter calculations. 
As it is known, saturation points obtained using different NN potentials with non-relativistic many-body approaches lie within a narrow 
band\ \cite{coester70,day81,ZHLi06}, the so-called Coester band, with either a too large saturation density or 
a too small binding energy compared to the empirical value.  Three-nucleon forces allows to reproduce properly the empirical saturation point. It has been pointed out that for similar reasons, three-body forces involving hyperons (NNY, NYY and YYY) may play also an important role to describe accurately the properties of neutron stars with hyperons \cite{taka02,taka08,isaac11,yama13,yama14,yama16,diego15} and hypernuclei \cite{spitzer,bach,dalitz,chalk63,gal66,gal67,lonardoni13,lonardoni14,cont18}.

 One of the longstanding open problems in nuclear physics and astrophysics that could be (if not completely at least partially) solved with the help of hyperonic three-body forces  (YTBF) is the so-called ``hyperon-puzzle" of neutron stars \cite{vida16,bombaci17}, {\it i.e.,} the difficulty to reconcile the measured masses of neutron stars with the presence of hyperons in their interiors.  Hyperons are expected to appear at $2-3$ times normal nuclear saturation density ($n_0=0.16$ fm$^{-3}$).  At such densities, the neutron and proton chemical potentials are large
enough to make the conversion of nucleons into hyperons energetically favorable. This conversion, however, produces a strong softening of the equation of state (EoS), due to the release of the Fermi pressure of the system, which leads to a decrease 
of the maximum neutron star mass predicted by theoretical models. In many microscopic calculations \cite{isaac00,baldo00,hans06,hans11,dapo10,riken16}, this decrease is so large that the maximum mass obtained is not compatible with the current largest measured neutron star masses  of $\sim 2M_\odot$ \cite{demo2010,arzo18,anto2013,cro19}.   
Most of these microscopic calculations have been performed using NN, NNN and NY interactions and, in some cases, also the YY 
one\ \cite{isaac00,riken16}. 
Just a few full consistent calculations including YTBF are present in literature. 
The authors of the present work, for instance, in Ref.\ \cite{isaac11} used a model based on the  Brueckner--Hartree--Fock (BHF) approach of hyperonic matter using the Argonne V18 \cite{av18} NN force and the Nijmegen NY soft- core NSC89 \cite{nsc89} one supplemented with additional simple phenomenological density-dependent contact terms to establish numerical lower and upper limits to the effect of the YTBF on the maximum mass of neutron stars. Assuming that the strength of these forces was either smaller than or as large as the pure nucleonic ones, the results of that work \cite{isaac11} showed that although the employed YTBF stiffened the EoS,  they were, however, unable to provide the repulsion needed to make the predicted maximum masses compatible with the recent observations of massive neutron stars. 
A multi-Pomeron exchange potential (MPP) model to introduce universal three-body repulsion among three baryons
in the hyperonic matter EoS was proposed in Refs.\  \cite{yama13,yama14,yama16}.
This universal three-body repulsive potential was based on the extended soft core (ESC) baryon-baryon interaction of the Nijmegen group \cite{esc06,esc06b}. The strength of the MPP was determined by analyzing the nucleus-nucleus scattering 
with the use of a G-matrix folding potential derived from the ESC interaction complemented with the MPP and a three-nucleon attractive part, added phenomenologically in order to reproduce the nuclear saturation properties. The results of those
works \cite{yama13,yama14,yama16} showed that when the MPP contribution was taken into account universally for all baryons, a maximum mass of  $\sim 2.2 M_\odot$ was obtained, in contradiction with the results and conclusions of Ref.\ \cite{isaac11} where the case of a universal three-body repulsion was also analyzed. Finally, in Ref.\ \cite{diego15} a Monte Carlo calculation of pure neutron matter with a non vanishing $\Lambda$-hyperon concentration was carried out 
including NN, NNN, N$\Lambda$ and NN$\Lambda$ forces. In particular the NN$\Lambda$ force used in that work was 
tuned in order to provide a reasonable description of the measured $\Lambda$ separation energy of several hypernuclei. The authors of Ref.\ \cite{diego15} concluded that, with the model they considered, the presence of hyperons in 
the core of neutron stars could not be satisfactory established and, consequently, according to these authors, there is no clear incompatibility with astrophysical observations when $\Lambda$s are included.
However, one should note, that the presence of protons, necessary to establish the correct  $\beta$-equilibrium inside neutron stars and thus a proper treatment of nuclear matter, was neglected in their calculation.
Although at present there is  not yet a general consensus regarding the role played by YTBF in the solution of the hyperon puzzle, it seems that even if they are not the full solution most probably they can contribute to it in an important way.

The aim of the present paper is to perform a calculation of the EoS and structure of neutron stars with non vanishing $\Lambda$-hyperon 
concentrations in the framework of non-relativistic  BHF approach (see {\it e.g.,} Ref.\ \cite{vidana00a,mythesis}) using realistic NN, NNN interactions derived in chiral effective field theory ($\chi$EFT) supplemented by  N$\Lambda$  and  NN$\Lambda$ interactions.
In particular, for the two-body NN interaction we use the local chiral potential presented in Ref.\ \cite{maria_local} at next-to-next-to-next-to-leading order (N3LO)
which includes the $\Delta(1232)$ isobar
 in the intermediate state of the NN scattering. Regarding the NNN force, we use the potential derived in Ref.\ \cite{N2LO} calculated at the 
next-to-next-to-leading-order (N2LO) of chiral perturbation theory in the local version as reported in Ref.\ \cite{N2LOL,logoteta16}. 
We note that in this NNN force the possibility of the $\Delta$-excitation is also taken into account. The low energy constants of the NNN have been fitted as in Ref.\ \cite{logoteta16} where it was shown that a good description 
of nuclear matter can be achieved using this setting. These interactions have been recently employed in Ref.\ \cite{BL} to calculate the 
$\beta$-stable EoS of nuclear matter and the structure of neutron stars. 
It was found a neutron star maximum mass of $2.07 M_\odot$ in agreement with the largest measured neutron star masses. The resulting EoS has been also recently used in 
Ref.\ \cite{endrizzi18} to simulate the merging of two equal mass neutron stars. 

In the present work we want to study how the 
finding of Ref.\ \cite{BL} changes allowing for the possible presence of $\Lambda$-hyperons in core of neutron stars. 
Although there is a vast number of NN and NNN interactions derived so far in $\chi$EFT, NY and YY interactions have been constructed only by the J\"{u}lich--Bonn--Munich group within this framework \cite{polinder06,haiden_ny,haiden_yy},
developing first the NY ones at leading order (LO) \cite{polinder06} and next-to-leading order (NLO) \cite{haiden_ny}, and then the YY constructed also at NLO \cite{haiden_yy}.  Since in the nucleonic sector we employ interactions calculated in $\chi$EFT, for consistency it would be appropriate to use hyperonic interactions derived in the same framework. 
Unfortunately, however, at present we do not have  at our disposal the N$\Lambda$ interaction presented in Refs.\ \cite{polinder06,haiden_ny} and, therefore, in this work we employ instead the N$\Lambda$ meson-exchange interaction derived by the Nijmegen group in Refs.\ \cite{nsc97a,nsc97}. We are aware that this represents the weakest point of this work that, however, we will try to solve in the future. 

Finally, concerning the YTBF, we use the NN$\Lambda$ force recently derived also by the J\"{u}lich--Bonn--Munich group in the framework of $\chi$EFT \cite{pesh_nny}.  In the next section we give some additional details on this force and how it is included in our BHF approach. The results of the calculation are then shown and discussed. Conclusions are given at the end.    

\begin{table*}
\begin{center}
\begin{tabular}{c|cccc}
\hline
\hline
               &  NSC97a+NN$\Lambda_1$  & NSC97a+NN$\Lambda_2$ & NSC97e+NN$\Lambda_1$ & NSC97e+NN$\Lambda_2$ \\
\hline               
               & $1.15$ & $1.68$ &  $1.39$ & $1.94$ \\ 
 \hline
 \hline
\end{tabular}
\caption{Values of the parameter $\beta$ for the different sets of N$\Lambda$ and NN$\Lambda$ interactions considered.}
\label{tab1}
\end{center}
\end{table*}


\section{The NN$\Lambda$ interaction}

As mentioned before, the construction of general three-baryon interactions within the framework of $\chi$EFT has been carried out by Petschauer {\it et al.,} in Ref.\ \cite{pesh_nny}.
The authors of this work have shown that the first contributions to NNY interactions in $\chi$EFT appear at N2LO.
They considered the contributions of three different classes of irreducible diagrams: three-baryon contact terms, one-meson exchange and two-meson exchange (see Fig.\ 1 of Ref.\ \cite{pesh_nny}). 
The pion-exchange mechanism is expected to be the dominant one while contributions coming 
from heavier meson exchanges (like $K$ or $\eta$ mesons) can be effectively absorbed into the contact terms. 
In addition, these authors in Ref.\ \cite{pesh_ny_eff} have derived an effective density dependent N$\Lambda$ interaction by averaging over the coordinates of one of the two nucleons. This allows a straightforward inclusion of the NN$\Lambda$ force in the BHF approach. 
The strategy is formally identical to the one adopted for the inclusion of the NNN interaction (see {\it e.g.,} Refs.\ \cite{baldo99,domenico15} for details). A more delicate point that deserves some comments concerns the setting of the low energy constants (LECs) in the NN$\Lambda$ interaction. In the pure nucleonic sector it is possible to use different choices for fixing the values of the LECs. Usually the LECs of the NNN interaction are fixed to reproduce the binding energy of light ($^3$H, $^3$He and $^4$He) nuclei but also other choices are possible. 
In the hypernuclear sector the situation is more 
complicated due to the lack of enough experimental data and to the few existing ab-initio calculations in light hypernuclei \cite{cont18,polinder06,haiden_ny,nogga12,ferrari17}.  
In Ref.\ \cite{pesh_ny_eff} in order estimate the values of the LECs, the baryon decouplet has been introduced as effective degree of freedom. Then a minimal non-relativistic Lagrangian has been constructed and the LECs have been estimated through decouplet saturation\ \cite{pesh_ny_eff}. 
In this approximation only one LEC, denoted as $H'$ in Ref.\ \cite{pesh_ny_eff}, remains as a free parameter. According to dimensional analysis in Ref.\ \cite{pesh_ny_eff} it has been considered $H'=\pm 1/f^2_{\pi}$ being $f_\pi=93$ ${\rm MeV}$ the pion decay constant. 
In the present, work we consider $H'= \beta/f^2_{\pi} $ where 
$\beta$ is a rescaling parameter that we fix in order to reproduce the single-particle potential at zero momentum, $U_\Lambda(0)$, of the $\Lambda$-hyperon in symmetric nuclear matter at saturation density. 
The empirical value of $U_\Lambda(0)$ is obtained from the extrapolation to infinite matter of the binding energy of the $\Lambda$ in hypernuclei, and it is found to be in the range $[-30, -28]$ MeV \cite{millener88}. In this work we consider the two extreme values of this interval to determine the parameter $\beta$. Hereafter we refer to NN$\Lambda_1$ and NN$\Lambda_2$  to the models in which $U_\Lambda(0)$ has been respectively taken equal to $-28$ MeV and $-30$ MeV to fix the value of $\beta$.
In order to regularize the short range part of the NN$\Lambda$ interaction, following\ \cite{pesh_ny_eff}, we have employed a non local regulator of the form $e^{-(p^4+p'^4)/\Lambda^4}$ with a  cut-off $\Lambda=500$ ${\rm MeV}$. Concerning the N$\Lambda$ interaction, as stated before, we have used the Nijmegen Soft-Core 97 (NSC97) meson-exchange NY interaction \cite{nsc97a,nsc97}. We note that the NSC97 NY force has been provided in $6$ different versions (NSC97a-f) according to the value of the magnetic vector $\alpha_V^m=F/(F+D)$ ratio. 
 For simplicity, in this work we have considered as representative cases of the N$\Lambda$ interaction the models NSC97a and NSC97e. Results for the other NSC97 models are qualitatively similar. Note that the parameter $\beta$, reported in Tab.\ \ref{tab1}, is in fact fixed for each set of N$\Lambda$ and NN$\Lambda$ interaction models. 
     
\section{Results and discussions}
\label{sec:results}

Before analyzing the effect of the NN$\Lambda$ interaction on neutron stars, it is interesting to consider first how this interaction affects the hypernuclear structure. To such end, we calculate the separation energy of the $\Lambda$ hyperon in a few hypernuclei.  This quantity is simply the difference between the total binding energies of an ordinary nucleus $^AZ$  and the corresponding hypernucleus $^{A+1}_\Lambda Z$. To determine it we follow a perturbative many-body approach to calculate the $\Lambda$ self-energy in finite nucleus which is then used to obtain, by solving the Schr\"{o}dinger equation, the energies and wave functions of all the single-particle bound states of the $\Lambda$ in the nucleus. A detail description of this method is given in Refs.\ \cite{mythesis,hyper0,hyper1,hyper2}. Results for the $\Lambda$ separation energy in $^{41}_\Lambda$Ca, $^{91}_\Lambda$Zr and $^{209}_\Lambda$Pb are shown in Tab.\ \ref{tab2}. We note that for technical reasons we have considered only hypernuclei that are described as a closed shell nuclear core plus a $\Lambda$ sitting in a single-particle state. Unfortunately, experimental data does not exists for the three hypernuclei considered and for comparison we have taken the closest representative ones for which experimental information is available. Data have been taken from Tab. IV of Ref.\ \cite{gal18}. Note that in the case of $^{91}_\Lambda$Zr and $^{209}_\Lambda$Pb hypernuclei, the inclusion of the NN$\Lambda$ interaction clearly improves the agreement of the theorerical calculation with the experimental data. This is, however, not the case of the  $^{41}_\Lambda$Ca where the two models of the NN$\Lambda$ interaction predict too much repulsion and none of them is able to provide a value of the $\Lambda$ separation energy  in agreement with the experimental one. The same trend has been observed in lighter hypernuclei. We recall, however, that all the finite hypernuclei results shown in Tab.\ \ref{tab2} have been obtained without any refitting of the parameter $\beta$, and that a better agreement with the experimental data for the lighter hypernuclei could in principle be obtained by readjusting this parameter individually to each hypernucleus. However,  a detailed study of the effect of the NN$\Lambda$ interaction on hypernuclei is out of the scope of the present work, and it is left for the future.

\begin{table} 
 \begin{center}
\begin{tabular}{l|ccc}
\hline
\hline
                &   $^{41}_{\Lambda}$Ca  &  $^{91}_{\Lambda}$Zr & $^{209}_{\Lambda}$Pb \\               
\hline
NSC97a                               &     $23.0$      &    $31.3$    &   $38.8$  \\  
NSC97a+NN$\Lambda_1$  &     $14.9$      &    $ 21.1$   &   $26.8 $    \\   
NSC97a+NN$\Lambda_2$  &     $13.3$      &    $19.3$    &   $ 24.7$  \\   
 \hline
NSC97e                               &     $24.2$       &    $32.3$    &   $39.5 $   \\ 
NSC97e+NN$\Lambda_1$ &      $ 16.1$      &    $22.3$    &  $27.9$   \\   
NSC97e+NN$\Lambda_2$ &      $14.7$       &    $20.7$    &  $26.1$  \\  
 \hline    
Exp.               &      $18.7(1.1)^\dagger$  &   $ 23.6(5)$    &   $26.9(8)$  \\
\hline
\hline
 \end{tabular}
\caption{$\Lambda$ separation energies (in MeV) of $^{41}_{\Lambda}$Ca, $^{91}_{\Lambda}$Zr and $^{209}_{\Lambda}$Pb for the different models considered with and without the inclusion of the NN$\Lambda$ force. Experimental results, taken from 
Tab. IV of Ref.\ \cite{gal18}, are shown for the closest measured hypernucleus
$^{40}_{\Lambda}$Ca, $^{89}_{\Lambda}$Y and $^{208}_{\Lambda}$Pb. $^\dagger$The weak signal for $^{40}_{\,\,\,\Lambda}$Ca \cite{Pile:1991} is not included in the recent compilation of Ref.\ \cite{gal18}.}
\label{tab2}
 \end{center}
\end{table} 
 
\begin{figure}
\begin{center}
\vspace{1cm}
\includegraphics[scale=0.3,angle=0]{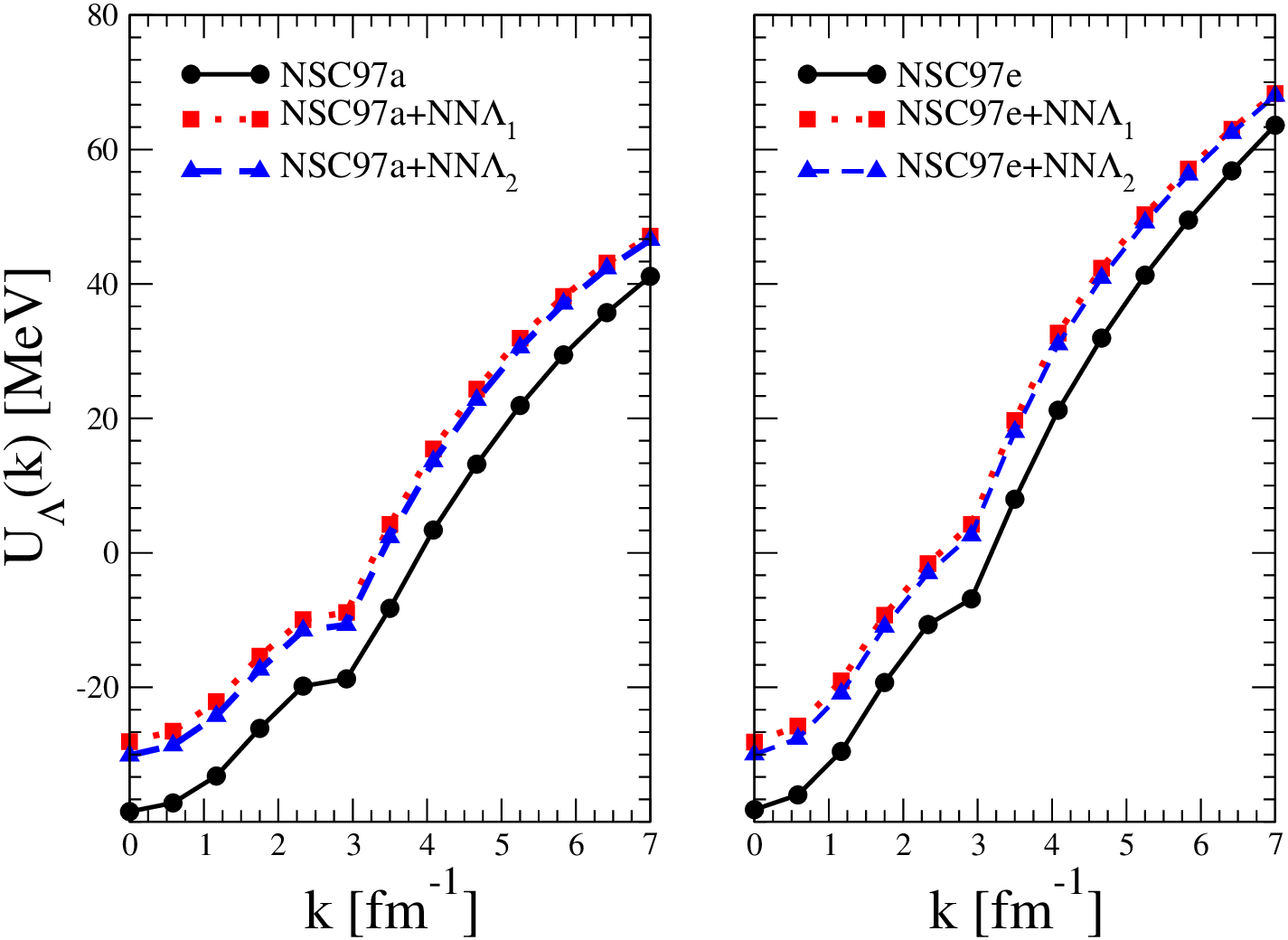}
\caption{(Color on-line) Single particle potentials for the $\Lambda$ hyperon in symmetric nuclear matter at saturation density ($n_0=0.16$ fm$^{-3}$) with and without the NN$\Lambda$ force.} 
\label{fig1}
\end{center}
\end{figure}

\begin{figure}
\begin{center}
\vspace{1cm}
\includegraphics[scale=0.3,angle=0]{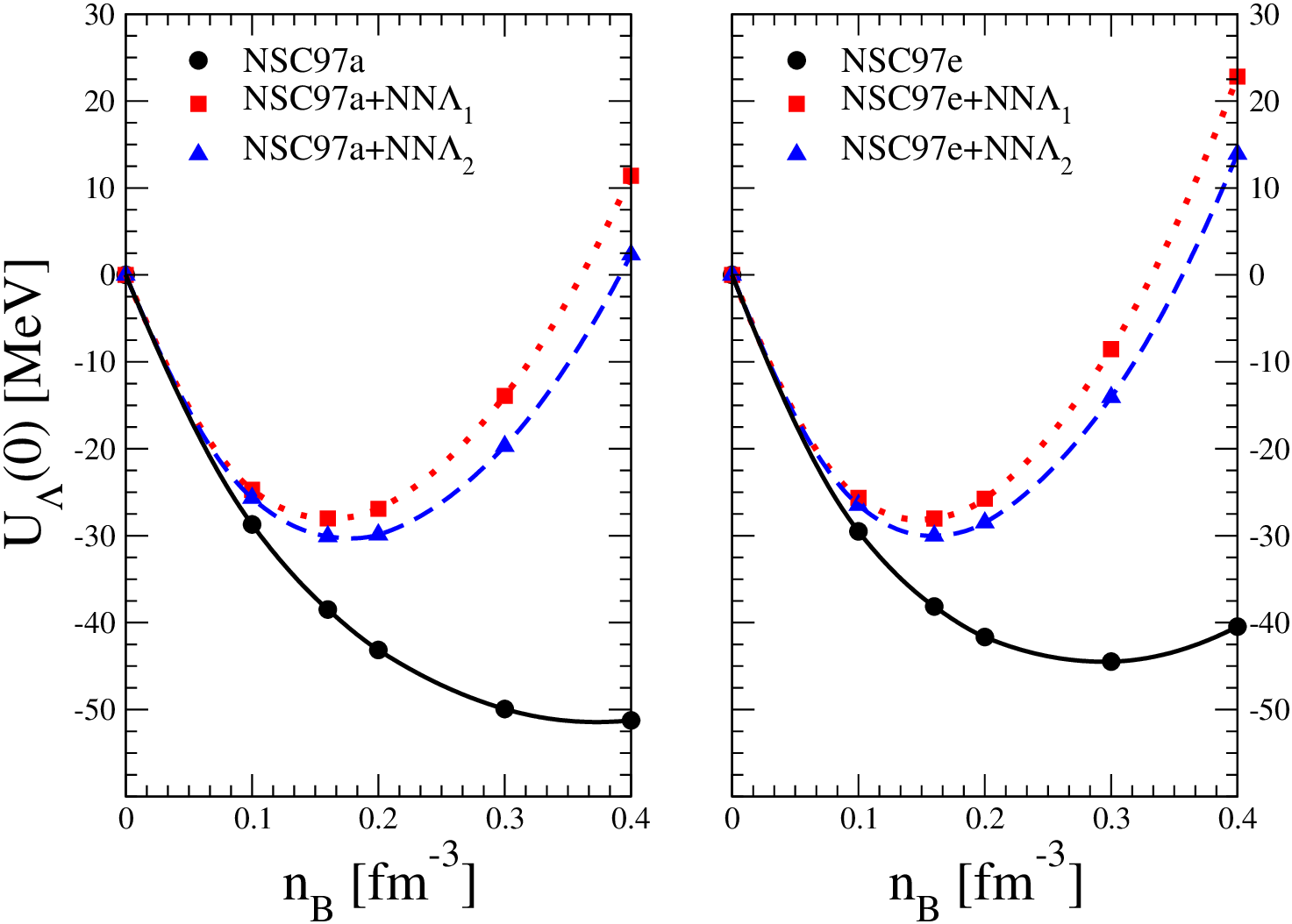}
\caption{(Color on-line) $U_\Lambda(0)$ as function of the baryonic density $n_B$ in symmetric nuclear matter with and without the NN$\Lambda$ force.} 
\label{fig2}
\end{center}
\end{figure}

Let us now consider the effect of the YTBF on neutron stars. In Fig.\ \ref{fig1} we show first the single particle potential, $U_\Lambda(k)$,  of the $\Lambda$-hyperon in symmetric nuclear matter as function of the single particle momentum $k$ at saturation density. Results for the NSC97a (NSC97e) N$\Lambda$ interaction are presented in the left (right) panel together with those including the effect of the NN$\Lambda$ force. Note that both the NSC97a and NSC97e models (with no $NN\Lambda$ interaction) predict a value of the $\Lambda$ single-particle potential at zero momentum of about $-40$ MeV, much lower than the empirical value extrapolated from hypernuclear data \cite{millener88}. Note also that the NSC97a model predicts more attraction than the NSC97e one over the whole range of momenta. This does not change adding the NN$\Lambda$ force. 
The repulsive effect of the NN$\Lambda$ interaction is even more clear looking at Fig.\ \ref{fig2} where $U_{\Lambda}(0)$ in symmetric nuclear matter is shown as function of the baryonic density $n_B$. Note that $U_\Lambda(0)$ is very deep when only two body interactions are considered and it shows a minimum located at $n_B\sim 0.4$ ${\rm fm}^{-3}$ and $n_B\sim 0.3$ ${\rm fm}^{-3}$ for the NSC97a and NSC97e models, respectively. 
The inclusion of the NN$\Lambda$ interaction induces repulsion for densities larger than about $0.1$ ${\rm fm}^{-3}$ and it shifts this minimum to a value of the density around $0.16$ ${\rm fm}^{-3}$ for all the models considered. As expected, the effect of YTBF is almost negligible in the low density region.

\begin{figure}
\begin{center}
\vspace{1cm}
\includegraphics[scale=0.3,angle=0]{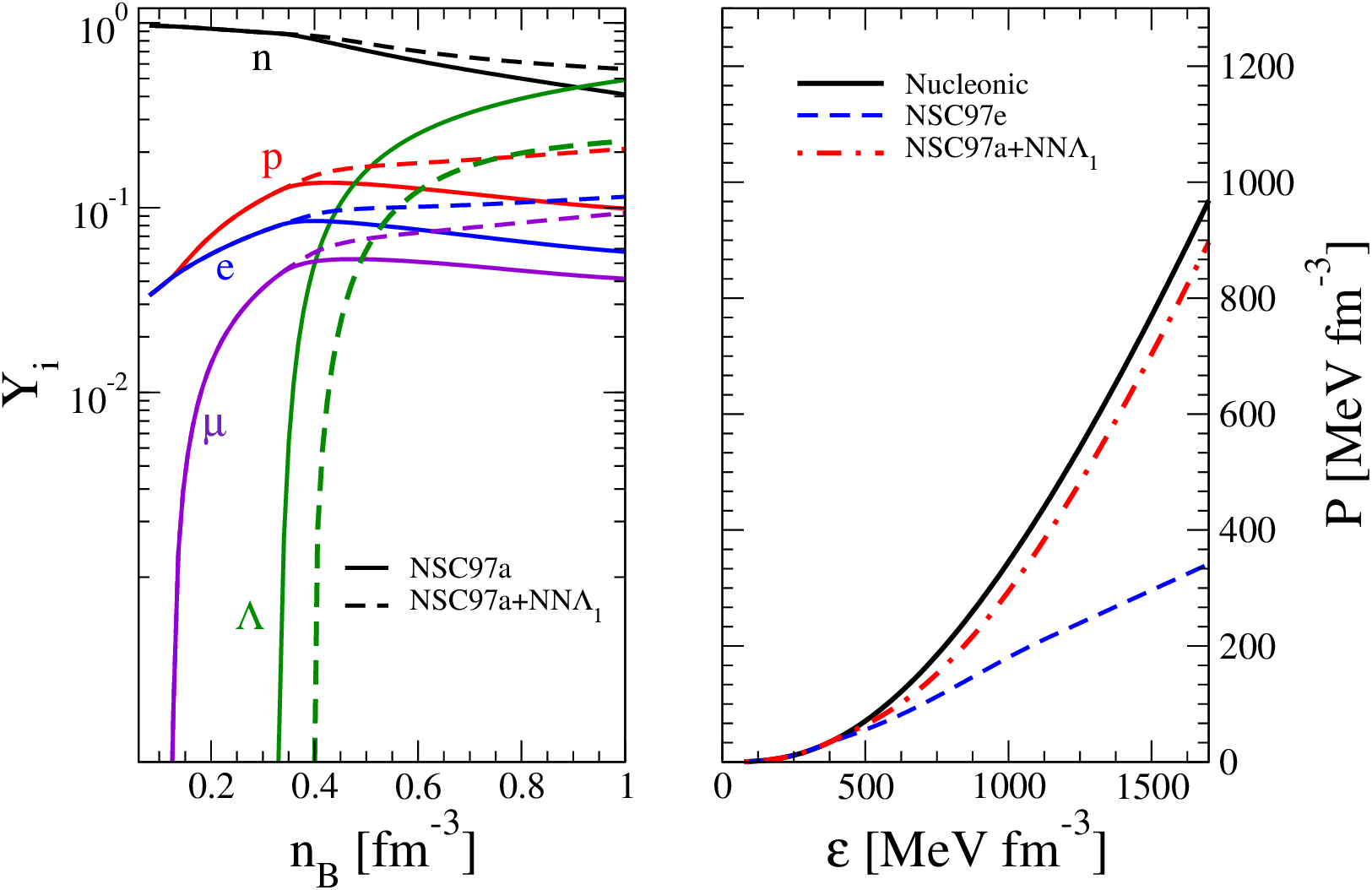}
\caption{(Color on-line) Composition (left panel) and EoS (right panel) of $\beta$-stable neutron star matter 
for models NSC97a (continuous lines) and NSC97a+NN$\Lambda_1$ (dashed lines). The EoS of
the pure nucleonic EoS is also shown for comparison.} 
\label{fig3}
\end{center}
\end{figure}

In order to perform the calculation of the $\beta$-stable neutron star matter EoS 
one has to find for each value of the total baryonic density $n_B=n_n+n_p+n_\Lambda$ the values of the particle concentration $Y_i=n_i/n_B$ 
that fulfill the chemical equilibrium equations: 
\begin{equation}
\mu_n-\mu_p=\mu_e, \,\,\,\,\,\,\,\,\, \mu_n=\mu_\Lambda, \,\,\,\,\,\,\,\, \mu_e=\mu_\mu . 
\end{equation}
Note that, besides nucleons and leptons, we have considered here only the $\Lambda$ and have ignored the possible appearance of other hyperons. The reason is that this is a first exploratory work where we are just interested on the role of the NN$\Lambda$ 
force. A more complete study of the effect of YTBF in neutron stars requieres, of course, the inclusion of the other hyperon species and their interactions. This, however, is  left for a future work.
In addition, the charge neutrality condition, $n_p=n_e+n_\mu$, should hold. In these equations $\mu_i$ and $n_i$ are, respectively, the chemical potential and number density of the $i$-th species. The chemical potential is calculated according to the usual thermodynamical relation: $\mu_i=\frac{\partial \epsilon}{\partial n_i}$ where $\epsilon$ is the energy density. 

\begin{figure}
\begin{center}
\vspace{1cm}
\includegraphics[scale=0.325,angle=0]{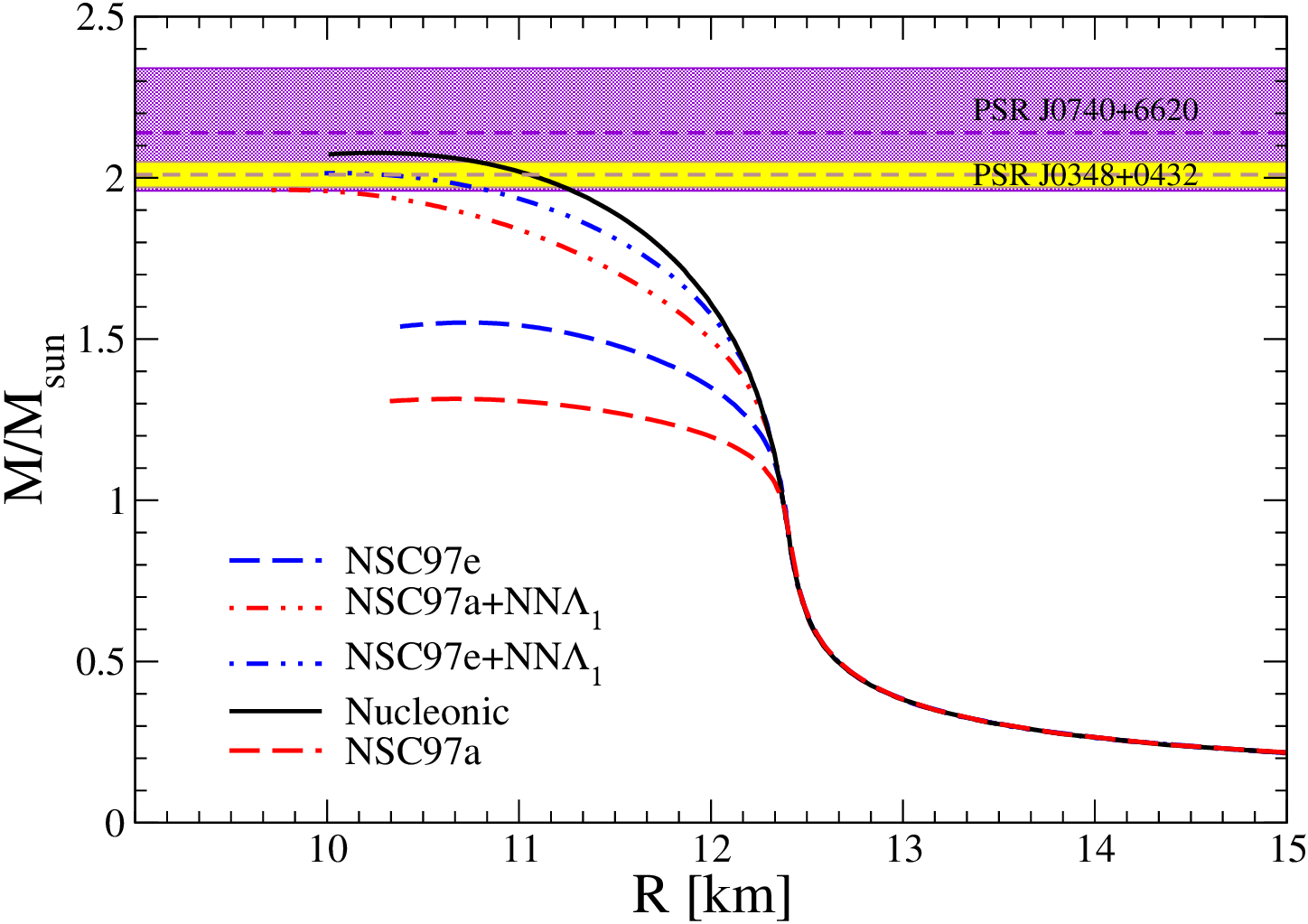}
\caption{(Color on-line) Mass-radius relation sequences for all the models considered. Results for pure nucleonic stars are shown for comparison. The  observed masses of the 
pulsars PSR J0348+0432 \cite{anto2013} and PSR J0740+6620 \cite{cro19} are also shown. The bands indicate the error of the observation.} 
\label{fig4}
\end{center}
\end{figure}

The composition of $\beta$-stable neutron star matter is shown in the left panel of Fig.\ \ref{fig3} for the models NSC97a  and NSC97a+ NN$\Lambda_1$. Qualitatively similar results are obtained for the other models which are not shown for simplicity. The continuous lines show the results when only N$\Lambda$, in addition to NN and NNN forces, are taken into account whereas the dashed ones include also the contribution of the NN$\Lambda$ force. The effect of the latter is twofold. First it shifts the onset of the $\Lambda$-hyperon to slightly larger baryonic densities. The second effect, maybe the most important one, is that the NN$\Lambda$ force strongly reduces the 
abundance of $\Lambda$ particles at large baryonic densities with the consequent stiffening of the EoS compared to the case in which the NN$\Lambda$ force is not included, as it can be seen in the right panel of the figure, where the total pressure $P$ is show as a function of the total energy density $\varepsilon$. Consequently, the mass of the neutron star, and in particular its maximum value, increases. This is shown in Fig.\ \ref{fig4} where it is plotted the mass-radius relation for the models NSC97a and NSC97e with and without the inclusion of the NN$\Lambda$ force obtained by solving the well known Tolman--Oppenheimer--Volkoff equations. The black line corresponds to the case of pure nucleonic matter shown as a reference. It is remarkable that the maximum masses obtained including the NN$\Lambda$ force are compatible with the largest measured masses of $\sim 2M_\odot$ \cite{demo2010,arzo18,anto2013,cro19}.   
This is in agreement with the calculation performed in Ref.\ \cite{diego15}. Notice that the result of our present calculations are based on a more realistic description of neutron star matter compared to the one given in Ref.\ \cite{diego15} (pure neutron matter plus a finite concentration of $\Lambda$ hyperons). In addition, we  
use more realistic interactions both in the nucleonic and the hyperonic sectors than the ones used in Ref.\ \cite{diego15}. Note also that, although the concentration of the $\Lambda$s is strongly reduced due to the effect of the NN$\Lambda$ force, they are still present in 
the interior of a $2 M_\odot$ neutron star. This differs from what is concluded in Ref.\ \cite{diego15} where it was found that the only NN$\Lambda$ force able to produce a EoS stiff enough to support maximum masses compatible with the recent observation of $2M_\odot$ neutron stars lead to the total disappearance of $\Lambda$ hyperons in the  core of these objects.

The neutron star properties, mass, radius and central baryonic density, for the maximum mass configuration are summarized in Tab.\ \ref{tab3}.  Note that models which do not account for the NN$\Lambda$ interaction provide very low neutron star maximum masses between $1.3-1.5M_\odot$. This is in agreement with several calculations performed by various research groups using different many-body methods \cite{isaac00,baldo00,hans06,hans11,dapo10}. 

\section{Conclusions}

We have studied the effects of a hyperonic NN$\Lambda$ force derived by the J\"{u}lich--Bonn--Munich in $\chi$EFT at  N2LO \cite{pesh_nny} in neutron stars and some single-$\Lambda$ hypernuclei. We have calculated the EoS and structure of neutron stars within the many-body BHF approach using in addition to the NN$\Lambda$ force realistic NN, NNN and N$\Lambda$ interactions. In particular, we have used the chiral NN and NNN interactions derived by Piarulli {\it et al.,} and Epelbaum {\it et al.} in 
Refs.\ \cite{maria_local} and \cite{N2LO}, respectively. For the N$\Lambda$, instead, we have employed the NSC97a and NSC97e models developed by the Nijmegen group within the framework of meson-exchange theory in Refs.\  \cite{nsc97a,nsc97}.
The reason for the use of this N$\Lambda$ interaction is simply the fact that we do not have presently at our disposal the chiral N$\Lambda$ interaction derived by the J\"{u}lich--Bonn--Munich group in Refs.\  \cite{polinder06,haiden_ny,haiden_yy}. This represents a weak point of the present work that, however, we will try to solve in the future. 
After adjusting the NN$\Lambda$ force to reproduce the binding energy of the $\Lambda$-hyperon in symmetric nuclear matter at saturation density, we have calculated the $\Lambda$ separation energy in $^{41}_\Lambda$Ca, $^{91}_\Lambda$Zr and
$^{209}_\Lambda$Pb. We have found that whereas the agreement between the calculated separated energy and the experimental data improves in the case of the heavier nuclei when the effect of the NN$\Lambda$ is included, this force results to be too much repulsive in the case of $^{41}_\Lambda$Ca and the lighter hypernuclei. We note, however, that all the finite hypenuclei results were obtained without refitting the NN$\Lambda$ force and that, a better agreement with experimental data for the lighter hypernuclei could be found if the force is adjusted individually to each hypernucleus. Finally, we have calculated the neutron star composition and EoS and have determined the maximum mass predicted by the different models considered. 
Our results have shown that when the NN$\Lambda$ force is included, the EoS becomes stiff enough such that the resulting maximum mass is compatible with largest measured neutron star maximum mass of $\sim 2M_\odot$. However, we have ignored the possible presence of other hyperon species in the neutron star interior that could change this conclusion, although, we should point out that hypothetical repulsive NNY, NYY and YYY  forces could lead to a similar one. Unfortunately, the lack of experimental information prevents currently any realistic attempt to estimate the effect of such forces.  More experimental efforts are, therefore, needed. In particular, new informations about the presence of hyperons inside the core of neutron stars may be provided in the future through the observation,  with the help of the new generation of gravitational wave detectors like the Einstein telescope \cite{ET1,ET2,ET3}, of signals emitted in the post-merger phase of  binary neutron stars coalescence \cite{abbott1,abbott2}.      

\begin{table} 
 \begin{center}
\begin{tabular}{l|ccc}
\hline
\hline
                 & $M_{max}(M_\odot)$ &   $R$ (km)    & $n_c$ (${\rm fm}^{-3}$)    \\               
\hline
Nucleonic        &      2.08         &   10.26       &   1.15            \\
\hline
NSC97a           &      1.31         &   10.60       &   1.40                      \\
NSC97a+NN$\Lambda_1$   &      1.96         &    9.80       &   1.30                      \\ 
NSC97a+NN$\Lambda_2$   &      1.97         &    9.87       &   1.28                      \\   
 \hline
NSC97e           &      1.54         &   10.81       &   1.18                         \\
NSC97e+NN$\Lambda_1$   &      2.01         &   10.10       &   1.20                      \\   
NSC97e+NN$\Lambda_2$   &      2.02         &   10.15       &   1.19                    \\   
\hline
\hline
 \end{tabular}
\caption{Neutron star properties, mass ($M_{max}$), radius ($R$) and central baryonic density ($n_c$), for the maximum mass configuration for
the different models considered. Results for a pure nucleonic star are shown for comparison.}
\label{tab3}
 \end{center}
\end{table} 

\section*{Acknowledgments}

The authors thank Avraham Gal for his useful comments.
This work has been supported by ``PHAROS: The multi-messenger physics and astrophysics of compact stars",  COST Action CA16214.



\begin{thebibliography}{200}

\bibitem{kievsky2008} A. Kievsky, S. Rosati, M. Viviani, L. E. Marcucci and L. Girlanda, J. Phys. G {\bf 35}, (2008) 063101.

\bibitem{coester70} F. Coester, S. Cohen, B. Day, and C. M. Vincent, Phys. Rev. C  {\bf 1}, (1970) 769. 

\bibitem{day81}     B. Day, Phys. Rev. Lett. {\bf 47}, (1981) 226.

\bibitem{ZHLi06}    Z. H. Li, U. Lombardo, H.-J. Schulze, W. Zuo, L. W. Chen, and H. R. Ma,  
                    Phys. Rev. C {\bf 74}, (2006) 047304.  

\bibitem{taka02} T. Takatsuka et al., Eur. Phys. J. A {\bf 13}, (2002) 213.

\bibitem{taka08} T. Takatsuka et al., Prog. Theor. Phys. Suppl. {\bf 174}, (2008) 80

\bibitem{isaac11} I. Vida\~na, D. Logoteta, C. Provid\^{e}ncia, A. Polls, and I. Bombaci, Eur. Phys. Lett.  {\bf 94}, (2011) 11002.

\bibitem{yama13} Y. Yamamoto, T. Furumoto, N. Yasutake, and Th, A, Rijken, Phys. Rev. C {\bf 88}, (2013) 022801.

\bibitem{yama14} Y. Yamamoto, T. Furumoto, N. Yasutake, and Th, A, Rijken, Phys. Rev. C {\bf 90}, (2014) 045805.

\bibitem{yama16} Y. Yamamoto, T. Furumoto, N. Yasutake, and Th, A, Rijken, Eur. Phys. J. A  {\bf 52}, (2016) 19.

\bibitem{diego15} D. Lonardoni, A. Lovato, S. Gandolfi, and F. Pederiva Phys. Rev. Lett. {\bf 114}, (2015) 092301.

\bibitem{spitzer} R. Spitzer, Phys. Rev. {\bf 110}, (1958) 1190.

\bibitem{bach} G. G. Bach, Nuovo Cimento {\bf XI}, (1959) 73.

\bibitem{dalitz} R. H. Dalitz, 9$^{th}$ Int. Ann. Conf. on High-Energy Physics, Academy of Sciences, USSR, Vol. I, (1960) 587.

\bibitem{chalk63} J. D. Chalk III and B. W. Downs, Phys. Rev. {\bf 132}, (1963) 2727.

\bibitem{gal66} A. Gal, Phys. Rev. {\bf 152}, (1966) 975.

\bibitem{gal67} A. Gal, Phys. Rev. Lett. {\bf 18}, (1967) 568.

\bibitem{lonardoni13} D. Lonardoni, S. Gandolfi, and F. Pederiva, Phys. Rev. C {\bf 87}, (2013) 041303(R).

\bibitem{lonardoni14} D. Lonardoni, F. Pederiva, and S. Gandolfi, Phys. Rev. C {\bf 89}, (2014) 014314.

\bibitem{cont18} L. Contessi, N. Barnea, and A. Gal, Phys. Rev. Lett. {\bf 121}, (2018) 102502.

\bibitem{vida16} D. Chatterjee and I. Vida\~na, Eur. Phys. J A. {\bf 52}, (2016) 29.

\bibitem{bombaci17} I. Bombaci, JPS Conf. Proc. {\bf 17}, (2017) 101002.

\bibitem{isaac00} I. Vida\~na, A. Polls, A. Ramos, L. Engvik, and M. Hjorth-Jensen, Phys. Rev. C {\bf 62}, (2000) 035801.

\bibitem{baldo00} M. Baldo, G. F. Burgio and H.-J. Schulze, Phys. Rev. C {\bf 61} (2000) 055801. 

\bibitem{hans06} H.-J. Schulze, A. Polls, A. Ramos, and I. Vida\~na, Phys. Rev. C {\bf 73}, (2006) 058801.

\bibitem{hans11} H.-J. Schulze and T. Riken Phys. Rev. C {\bf 84}, (2011) 035801.

\bibitem{dapo10} H. Dapo, B.-J. Schaefer and J. Wambach Phys. Rev. C {\bf 81}, (2010) 035803. 

\bibitem{riken16} T. Rijken and H.-J. Shulze,  EPJ {\bf A52} (2016) 21. 

\bibitem{demo2010} P. Demorest, T. Pennucci, S. Ransom, M. Roberts, J. Hessels, Nature {\bf 467}, (2010) 1081.  

\bibitem{arzo18} Z. Azoumanian {\it et al.,} Astrophys. J. Suppl. {\bf 235}, (2018) 37.

\bibitem{anto2013} J. Antoniadis et al., Science {\bf 340}, (2013) 1233232.   

\bibitem{cro19} H. T. Cromartie {\it et al.,} Nature Astronomy {\bf 10.1038} (2019).

\bibitem{av18} R. B. Wiringa, V. G. J. Stoks, and R. Schiavilla, Phys. Rev. C {\bf 51}, (1995) 38.

\bibitem{nsc89} P. M. M. Maesen. T. A. Rijken, and J. J. de Swart, Phys. Rev. C {\bf 40}, (1989) 2226.

\bibitem{esc06} T. A. Rijken, Phys. Rev. C {\bf 73}, (2006) 044007.

\bibitem{esc06b} T. A. Rijken and Y. Yamamoto, Phys. Rev. C {\bf 73}, (2006) 04408.

\bibitem{vidana00a}  I. Vida\~na, A. Polls, A. Ramos, M. Hjorth-Jensen, and V. G. J. Stoks, Phys. Rev. C {\bf 61}, (2000) 025802.

\bibitem{mythesis} I. Vida\~na, Ph.D. thesis, Univesity of Barcelona, 2001, https://www.tesisenred.net/handle/10803/1583

\bibitem{maria_local} M. Piarulli, L. Girlanda, R. Schiavilla, A. Kievsky, A. Lovato, L. E. Marcucci, 
 S. C. Pieper, M. Viviani and R. B. Wiringa, Phys. Rev. C {\bf 94}, (2016) 054007.    

\bibitem{N2LO} E. Epelbaum, A. Nogga, W. Gl\"ockle, H. Kamada, and Ulf-G. Mei{\ss}ner, and H. Wita\l a,  
                               Phys. Rev. C {\bf 66}, (2002) 064001.


\bibitem{N2LOL} P. Navratil, Few-Body Syst. {\bf 41}, (2007) 117.

\bibitem{logoteta16} D. Logoteta, I. Bombaci and A. Kievsky, Phys Rev. C, {\bf 94} (2016) 064001.   

\bibitem{BL} I. Bombaci and D. Logoteta A\&A,  {\bf 609} (2018) A128. 

\bibitem{endrizzi18} A. Endrizzi, D. Logoteta, B. Giacomazzo, I. Bombaci, W. Kastaun and R. Ciolfi, Phys. Rev. D {\bf 98},  (2018) 043015. 

\bibitem{polinder06} H. Polinder, J. Haidenbauer, and U.-G. Mei{\ss}ner, Nucl. Phys. A {\bf 779}, (2006) 244.

\bibitem{haiden_ny} J. Haidenbauer, S. Petschauer, N. Kaiser, U.-G. Mei{\ss}ner, A. Nogga and W. Weise  
 Nucl. Phys. A {\bf 915} (2013) 24. 

\bibitem{haiden_yy} J. Haidenbauer, S. Petschauer and U.-G. Mei{\ss}ner, Nucl. Phys. A, {\bf 954}, (2016) 273.   

\bibitem{nsc97a} Th. A. Rijken, V. G. J. Stoks and Y. Yamamoto, Phys. Rev. C {\bf 59} (1999) 21.

\bibitem{nsc97}  V. G. J. Stoks, Th. A. Rijken, Phys. Rev. C {\bf 59} (1999) 3009.

\bibitem{pesh_nny} S. Petschauer, J. Haidenbauer, U.-G. Mei{\ss}ner and W. Weise, Phys. Rev. C {\bf 93} (2016) 014001. 

\bibitem{pesh_ny_eff} S. Petschauer, J. Haidenbauer, N. Kaiser, U.-G. Mei{\ss}ner and W. Weise, Nucl. Phys. A {\bf 957} (2017) 347.

\bibitem{baldo99} M. Baldo and L. S. Ferreira, Phys. Rev. C {\bf 59} (1999) 682.

\bibitem{domenico15} D. Logoteta, I. Vida\~na, I. Bombaci, and A. Kievsky, Phys. Rev. C {\bf 91}, (2015) 064001.

\bibitem{nogga12} A. Nogga, H. Kamada, W. G\"{o}ckle, Phys. Rev. Lett. {\bf 88}, (2012) 172501.

\bibitem{ferrari17} F. Ferrari Ruffino, N. Barnea, S. Deflorian, W. Leidemann, D. Lonardoni, G. Orlandina, and F. Pederiva, Few-Body Syst. {\bf 58}, (2017) 113. 

\bibitem{millener88} D. J. Millener, C. B. Dover and A. Gal, Phys. Rev. C {\bf 38} (1988) 2700.

\bibitem{hyper0} M. Hjorth-Jensen, A. Polls, A. Ramos, and H. M\"{u}ther, Nucl. Phys. A {\bf 605}, (1996) 458

\bibitem{hyper1} I. Vida\~na, A. Polls, A. Ramos and M. Hjorth-Jensen, Nucl. Phys. A {\bf 644} (1998) 201.

\bibitem{hyper2} I. Vida\~na, Nucl. Phys. A {\bf 958}, (2017) 48.

\bibitem{gal18} A. Gal, E. V. Hugenford, and D. J. Millener, Rev. Mod. Phys. {\bf 88} (2016) 035004.

\bibitem{Pile:1991} P.~H.~Pile {\it et al.},   Phys.\ Rev.\ Lett.\  {\bf 66}, 2585 (1991).

\bibitem{ET1} B. S. Sathyaprakash {\it et al.,} arXiv:1108.1423v2 (2012).

\bibitem{ET2} D. Meacher {\it et al.,} Phys. Rev. D {\bf 93}, (2016) 024018.

\bibitem{ET3} B. S. Sathyapraksh {\it et al.,} arXiv:1903.09221 (2019).

\bibitem{abbott1} B. P. Abbott et al., Phys. Rev. Lett. {\bf 119}, (2017) 161101.  

\bibitem{abbott2} B. P. Abbott et al., Astrophys. J. {\bf 848}, (2017) L13. 




\end{thebibliography}
\end{document}